\documentclass[aps,pra,reprint,showpacs,superscriptaddress,nofootinbib,onecolumn]{revtex4}
\usepackage{amsmath}
\usepackage{amsfonts}
\usepackage{amsthm}
\usepackage{amssymb}
\usepackage{graphicx}
\usepackage{enumerate}
\usepackage{color}
\usepackage[T1]{fontenc}
\usepackage{mathptmx}
\usepackage{braket}
\usepackage{todonotes}
\usepackage{soul}
\usepackage{subfigure}
\graphicspath{{figures/}}
\usepackage{booktabs}

\usepackage[bookmarks=false,colorlinks,citecolor=blue,
linkcolor=blue,anchorcolor=blue,urlcolor=blue
]{hyperref}

\begin{document}

\title{Open-destination measurement-device-independent quantum key distribution network}

\author{Wen-Fei Cao}
\author{Yi-Zheng Zhen}
\author{Yu-Lin Zheng}
\author{Shuai Zhao}
\author{Feihu Xu}
\email{feihuxu@ustc.edu.cn}
\author{Li Li}
\email{eidos@ustc.edu.cn}
\affiliation{Hefei National Laboratory for Physical Sciences at Microscale and Department of Modern Physics, University of Science and Technology of China, Hefei, Anhui 230026, People's Republic of China\\}
\affiliation{CAS Center for Excellence and Synergetic Innovation Center of Quantum Information and Quantum Physics, University of Science and Technology of China, Hefei, Anhui 230026, People's Republic of China\\}
\author{Zeng-Bing Chen}
\email{zbchen@nju.edu.cn}
\affiliation{National Laboratory of Solid State Microstructures and School of Physics, Nanjing University, Nanjing~210093, People's Republic of China\\}
\author{Nai-Le Liu}
\email{nlliu@ustc.edu.cn}
\author{Kai Chen}
\email{kaichen@ustc.edu.cn}
\affiliation{Hefei National Laboratory for Physical Sciences at Microscale and Department of Modern Physics, University of Science and Technology of China, Hefei, Anhui 230026, People's Republic of China\\}
\affiliation{CAS Center for Excellence and Synergetic Innovation Center of Quantum Information and Quantum Physics, University of Science and Technology of China, Hefei, Anhui 230026, People's Republic of China\\}


\begin{abstract}
Quantum key distribution (QKD) networks hold promise for sharing secure randomness over multi-partities.
Most existing QKD network schemes and demonstrations are based on trusted relays or limited to point-to-point scenario.
Here, we propose a flexible and extensible scheme named as open-destination measurement-device-independent QKD network.
The scheme enjoys security against untrusted relays and all detector side-channel attacks.
Particularly, any users can accomplish key distribution under assistance of others in the network. As an illustration,
we show in detail a four-user network where two users establish secure communication and present realistic simulations
by taking into account imperfections of both sources and detectors.
\end{abstract}


\maketitle


\section{Introduction}

Quantum key distribution (QKD)~\cite{Bennett1984, Ekert1991, Gisin2002, Scarani2009} provides unconditional security between distant communication parties based on the fundamental laws of quantum physics.
In the last three decades, QKD has achieved tremendous progress in both theoretical developments and experimental demonstrations.
To extend to a large scale, the~QKD network holds promise to establish an unconditionally secure global network.
Different topologies for QKD network have been demonstrated experimentally during the past decades~\cite{Elliott2005, Peev2009, Chen2010, Sasaki2011, Froehlich2013, Qiu2014, Liao2018}. However, due to high demanding on security and the relatively low detection efficiency, the~realization of large-scale QKD networks is still~challenging.

On the one hand, many previous demonstrations of quantum networks heavily rely on the assumption of trusted measurement devices.
From security point of view, however, such assumption is challenging in realistic situations, as~various kinds of detector side-channel attacks are found due to the imperfections of practical devices~\cite{Qi2007, Zhao2008, Lydersen2010, Gerhardt2011, Weier2011}.
Fortunately, measurement-device-independent QKD (MDI-QKD) protocol~\cite{Lo2012,Braunstein2012}
can remove all kinds of attacks in the detector side-channel.
Since its security does not rely on any assumptions on measurement devices, MDI-QKD networks are expected to close the security loophole existing in the previous QKD networks.
The MDI-QKD network has been discussed theoretically in Ref.~\cite{Xu2015c, Pirandola2015}, and~a preliminary experimental MDI-QKD network demonstration was realized very recently~\cite{Tang2016a}.

On the other hand, most of the existing QKD networks are limited to point-to-point QKD.
When~expanded to multi-partite QKD case, the~complexity increases, and~the efficiency decreases significantly.
Recent study shows that multi-partite entanglement can speed up QKD in networks~\cite{Epping2017}.
Therefore, it is highly desirable to develop variously novel schemes of QKD networks if assisted by multi-partite entanglement source.
Then, an~immediate problem comes out: how to design a QKD network enjoying security against untrusted measurement devices and simultaneously offer practical applicability for arbitrary scalability? This is exactly the purpose of this~work.

In this paper, we propose a flexible and extensible protocol named as open-destination MDI-QKD network, by~combining the idea of open-destination teleportation
~\cite{Zhao2004} and MDI-QKD~\cite{Lo2012,Braunstein2012}.
In~this protocol, secure communication between any two users in the network can be accomplished under assistance of others.
The open-destination feature allows these two-party users share secure keys simultaneously, where we also generalize to the case of $C$ communication users.
Remarkably, this~feature allows communication users not to be specified before the measurement step, which~makes the network flexible and extendable.
Furthermore, the~MDI feature enables this scheme to be secure against untrusted relays and all detector side-channel attacks.
Specially, all users need only trusted state-preparation devices at hand, while the untrusted relay section is made by entangled resources and measurement~devices.

\section{Open-Destination MDI-QKD~Network}
\label{sec:N_2_case}
Consider an $N$-party quantum network.
We are particularly interested in the case where arbitrary two users want to share secure keys.
This scenario is denoted as $(N,2)$ for convenience.
To simplify the discussion, here we focus on the star-type network, where both the user and a central source emit quantum signals.
The signals are measured by untrusted relays located between each user and the central~source.


\subsection{Protocol}

\begin{figure}[!htb]
\centering
\includegraphics[width=.7\columnwidth]{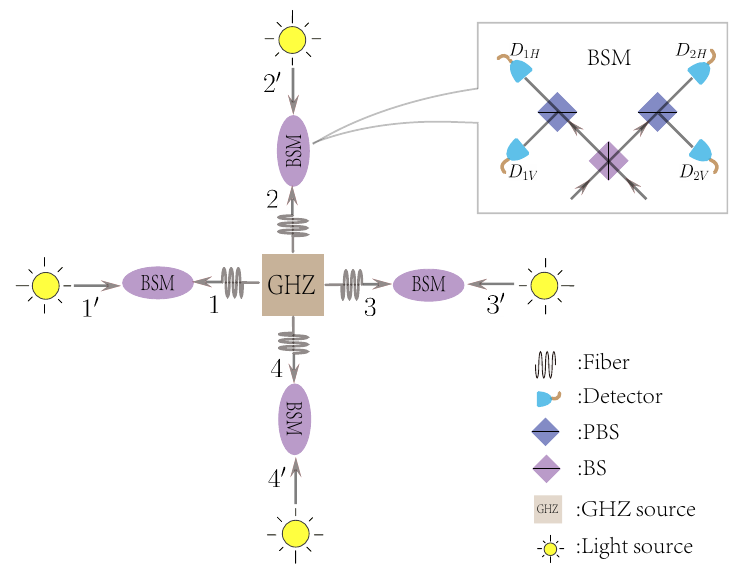}
\caption{
{An optical diagram for the polarization-encoding $(4,2)$ open-destination measurement-device-independent quantum key distribution (MDI-QKD) network.}~The GHZ source outputs $4$-partite GHZ entangled state in polarization and the light source outputs BB84 polarization state.
The BSM represents the Bell state measurement, where BS is the 50:50 beam splitter, PBS is the polarization beam splitter, and~$D_{1H}$, $D_{2H}$, $D_{1V}$, and~$D_{2V}$ are single-photon detectors.
A click in $D_{1H}$ and $D_{2V}$, or~in $D_{1V}$ and $D_{2H}$, indicates a projection into the Bell state $\ket{\psi^{-}}=(\ket{01}-\ket{10})/\sqrt{2}$, and~a click in $D_{1H}$ and $D_{1V}$, or~in $D_{2H}$ and $D_{2V}$, indicates a projection into the Bell state $\ket{\psi^{+}}=(\ket{01}+\ket{10})/\sqrt{2}$.
}
\label{fig:optical-diagram}
\end{figure}

The $(N,2)$ open-destination MDI-QKD runs as follows. An~illustration of the $(4,2)$ example is shown in Figure~\ref{fig:optical-diagram}.


\begin{enumerate}[labelsep=0]

\item[Step. 1] \textbf{Preparation}: A third party, which may be untrusted, prepares $N$-partite GHZ state
\begin{equation}
\ket{GHZ}_{N} = \frac{1}{\sqrt{2}}(\ket{0}^{\otimes N}+\ket{1}^{\otimes N}),
\end{equation}
where $\ket{0}$ and $\ket{1}$ denote two eigenstates of the computational basis $Z$.
All users prepare BB84 polarization states, i.e.,~$\ket{0}$, $\ket{1}$, $\ket{+}$, and~$\ket{-}$ with $\ket{\pm}= (\ket{0}\pm\ket{1})/\sqrt{2}$ being the two eigenstates of the
basis $X$.
The third party and all users distribute the prepared quantum states to their relays, which may also be~untrusted.

\item[Step. 2] \textbf{Measurement}: The relays perform Bell state measurements (BSMs).
When using linear optical setups, only two outcomes related to projections on $\ket{\psi^{\pm}}=(\ket{01}\pm\ket{10})/\sqrt{2}$ can be distinguished.

\item[Step. 3] \textbf{Announcement}:
All relays announce their successful BSM results among a public classical authenticated channel. The~two communication users announce their photons bases, and~other users announce their states prepared in the $X$ basis.

\item[Step. 4] \textbf{Sifting}: The two communication user keep the strings where all the relays get successful BSM results and other users use $X$ bases.
Then, they discard the strings where different preparation bases are used.
To guarantee their strings to be correctly correlated, one of the two users flip or not flip his/her bit according to the corresponding BSM results and other users' prepared states
(see Appendix~\ref{sec:sifiting} for details).
Then, the~two users obtain the raw key~bits.

\item[Step. 5] \textbf{Post-processing}:
The two communication users estimate the quantum phase error and quantum bit error rate (QBER) in $Z$ and $X$ bases, 
according to which they further perform error correction and privacy amplification to extract correct and secure keys.
\end{enumerate}

In this protocol, the~multi-partite GHZ state between distant users can also be established through a prior distributed singlets, following the scheme of Bose
\emph{et al.}~\cite{Bose1998}.
In fact, the~open-destination feature allows arbitrary two users in the network to share secure keys based on the same experiment statistics.
To accomplish the task of MDI-QKD among arbitrary two users, a~natural scheme is to establish direct MDI-QKD between each two users.
This requires either the central source to adjust his devices such that EPR 
pairs (the maximally entangled quantum states of a two qubit system, named after Einsetin, Podolski and Rosen Paradox~\cite{Einsetin1935}) are sent along desired directions, or~a number $N(N-1)/2$ of two-user combinations to establish direct MDI-QKD using the same number of untrusted relays.
The open-destination scheme is an alternative scheme.
It does not require the central source to adjust his devices according to the demand of communications, at~the same time involve only $N$ untrusted relays.
In a practical scenario, all the users can use weak coherent pulses to reduce experimental cost and apply decoy-state techniques~\cite{Hwang2003,Lo2005,Wang2005} to avoid photon-number-splitting attack, as~well as to estimate the gain and the error rate.


\subsection{Correctness and Security~Analysis}
We will show the correctness and security of the open-destination MDI-QKD protocol, i.e.,~the~communication users end up with sharing a common key in an honest run and any eavesdropper can only obtain limited information of the final key. The~following analysis applies for the $(N,2)$ case. As~an illustration, we show a detailed derivation of the $(4,2)$ in Appendix~\ref{sec:sifiting}.


For the correctness of the protocol, we show that after successful BSMs and other users announce the $X$-basis states, the~two communication users can perform flip their bits locally to obtain perfectly correlated sifted keys.
We start from rewriting the GHZ state as
\begin{align}
\ket{GHZ}_{N}
= & \frac{1}{\sqrt{2}}\left[
\ket{00}_{12}\bigotimes_{k=3\dots N}\frac{\ket{+}_k+\ket{-}_k}{\sqrt{2}}
+ \ket{11}_{12}\bigotimes_{k=3\dots N}\frac{\ket{+}_k-\ket{-}_k}{\sqrt{2}}
\right],\\
= & \left(\frac{1}{\sqrt{2}}\right)^{N-1}\sum_{\chi}
\left(\ket{00}_{12}+(-1)^{\sigma_\chi}\ket{11}_{12}\right)\ket{\chi}_{3\dots N}.
\label{eq:ghz-reformulating}
\end{align}

Here, $\chi\in\{+,-\}^{N-2}$ is a string of $N-2$ bits with bit value ``$+$'' or ``$-$'' and $\sigma_\chi=0(1)$ if the number of ``$-$'' is even (odd).

We label each user by $1^\prime,2^\prime,\dots,N^\prime$ and let the two communication users be $1^\prime$ and $2^\prime$.
In a successful run of the protocol, suppose that users $1^\prime$ and $2^\prime$ prepare states $\ket{\alpha},\ket{\beta}\in\{0,1,+,-\}$, respectively, and~other users $3^\prime,\dots,N^\prime$ prepare state in the $X$ basis, denoted as a string $\chi^\prime\in\{+,-\}^{N-2}$.
In addition, denote the successful BSM results as a string $\upsilon\in\{+,-\}^N$, with~the $k$th bit $\upsilon_k$ denoting the BSM outcome on the state prepared by the user $k^\prime$ and the $k$-th particle of the GHZ state.
Here, $\upsilon_k=\pm$ corresponds to projections $\ket{\psi^\pm}\bra{\psi^\pm}$, respectively.
Then, when other users send states denoted by $\ket{\chi^\prime}$ and when all untrusted relays announce successful BSM results $\upsilon$, the~equivalent measurement $M_{12}^{\chi^\prime,\upsilon}$ on $1^\prime$ and $2^\prime$ is
\begin{align}
\sqrt{M_{1^\prime 2^\prime}^{\chi^\prime,\upsilon}}\ket{\alpha\beta}_{1^\prime 2^\prime} =&
\left(\bigotimes_{k}\bra{\psi^{\upsilon_k}}_{kk^\prime}\right)\ket{GHZ}_N\otimes\ket{\alpha\beta}_{1^\prime 2^\prime}\ket{\chi^\prime}_{3^\prime \dots N^\prime},\\
=& \left(\frac{1}{\sqrt{2}}\right)^{N-1}\sum_{\chi}\bra{\psi^{\upsilon_1}}_{11^\prime}\bra{\psi^{\upsilon_2}}_{22^\prime}
\left(\ket{00}_{12}+(-1)^{\sigma_\chi}\ket{11}_{12}\right)\ket{\alpha\beta}_{1^\prime 2^\prime} \nonumber \\
&\qquad \times
\prod_{k=3\dots N}\bra{\psi^{\upsilon_k}}_{kk^\prime}\ket{\chi}_{k}\ket{\chi^\prime}_{k^\prime},\\
\propto &
\left(\bra{00}_{1^\prime 2^\prime} + (-1)^{\tau}\bra{11}_{1^\prime 2^\prime}\right)\ket{\alpha\beta}_{1^\prime 2^\prime}.
\end{align}

Here, $\tau=\sigma_{\chi^\prime\oplus\tilde{\upsilon}}\oplus\upsilon_1\oplus\upsilon_2$ with $\tilde{\upsilon}=\upsilon_3\upsilon_4\dots\upsilon_N\in\{+,-\}^{N-2}$ and $\sigma_{\chi^\prime\oplus\tilde{\upsilon}}=+(-)$ if the number of ``$-$'' in $\chi^\prime\oplus\tilde{\upsilon}$ is even (odd). Therefore, when the user $1^\prime$ and $2^\prime$ both prepare $Z$-basis states, or~when they both prepare $X$-basis states with $\tau=0$, the~corresponding strings are correctly correlated; otherwise, when they both prepare $X$-basis states but $\tau=1$, their strings are anticorrelated, and~one party needs to flip all his/her~bits.

For the security of the protocol, here we show
that an open-destination MDI-QKD can be equivalent to a standard bipartite MDI-QKD if we only focus on the two communication users.
Recall that, in~the standard MDI-QKD, two parties, Alice and Bob, prepare and send quantum signals to a remote untrusted relay, which announces a successful BSM result or not.
In our scheme,
one~can treat all parts outside the two users $1^\prime$ and $2^\prime$ as an untrusted relay~\cite{Xu2013a}.
That is, the~GHZ source, the~BSM setups and all other users serve as a big untrusted relay, and~the successful BSM results in the standard MDI-QKD corresponds to all BSMs announcing successful measurements together with all other users announcing $X$-basis states
(see Figure~\ref{fig:odqkd-security-proof-frame} as an example of the $(4,2)$ case).
In this sense, our scheme is reduced to the MDI-QKD and the two has the same security.
Additionally, although~we require the preparation device of each user to be trusted in the protocol, the two communication users need not to
trust these preparation devices of other users.


\subsection{Key Generation~Rate}
The key generation rate for open-destination MDI-QKD can be derived similarly as the standard MDI-QKD, i.e.,
by converting it to an entanglement purification scheme.
Suppose that the two communication users both have virtual singlets at their hands and then send one particle to the untrusted relays.
In a successful run of the protocol, the~remaining virtual particles of the two communication users will be entangled.
When the entanglement between the virtual particles is sufficiently strong, the~monogamy property of entanglement~\cite{Koashi2004, Osborne2006, Ou2008} guarantees the extraction of
information-theoretically secure key bits between the two users.
In this sense, the~secret key rate can be roughly viewed as the gains of entanglement purification in the asymptotic case.
Taking account of imperfections, such as basis misalignment, channel loss, and~dark counts of the detectors, the~key generation rate is given by the GLLP 
method~\cite{Gottesman2004}
\begin{equation}
R_2 = Q^{ZZ}\left[1 - H\left(e^{XX}\right)- f H\left(e^{ZZ}\right)\right].
\end{equation}

Here, we have assumed that the user $1^\prime$ and $2^\prime$ use $Z$ basis to generate keys and use $X$ basis to estimate phase errors.
In the equation, $Q^{ZZ}$ denotes the overall gain in the $Z$ basis, and~$e^{XX}$ ($e^{ZZ}$) denotes the phase (bit) error rate,
$f>1$ is the error correction inefficiency for the error correction process,
and $H(x) = -x \log_2 (x)-(1-x) \log_2(1-x)$ is the binary Shannon entropy function.
In a realistic experiment, if~using weak coherent pulses and adopting decoy-state techniques,
$Q^{ZZ}$, $e^{ZZ}$, and~$e^{XX}$
can be efficiently estimated~\cite{Lo2005,Wang2005}.

\subsection{Comparison with the Standard~MDI-QKD}

The~open-destination MDI-QKD network is different from the conventional MDI-QKD.
The~main difference comes from the open-destination feature, which in fact allows the all 2-party users in the network generate their own secure keys independently and simultaneously.
There are in fact \mbox{$N(N-1)/2$} combinations of such two-party users.
If one uses the conventional MDI-QKD scheme, the~same number of untrusted relays are required.
To increase the communication distance, one~may further add the same number of relays and EPR sources to construct the user-relay-EPR source-relay-user structure.
Such construction of quantum network could be expensive considering the number of devices required.
One could also use the optical switches to reduce the number of relays; however, in~this case the communication would be arranged in time order and some users have to wait.
In the open-destination scheme, $N$ untrusted relays are sufficient to connect each other supplied with good-quality GHZ central source.
Although the distribution of GHZ states may lead to other technological challenges, the~open-destination scheme can reduce the number of devices significantly in constructing the network.
As for the performance, the~two schemes in fact have similar performance in the ideal case.
The difference is that the open-destination scheme generates secure keys for any two-party users in one round of implementation while the bipartite MDI-QKD scheme costs $N(N-1)/2$ rounds.
Furthermore, the~open-destination scheme also establishes conference key agreements among arbitrary users, which can not be accomplished directly via the bipartite MDI-QKD.
We will discuss this case in the next~section.

\section{Numerical~Simulation}\label{sec:open-destination-ex}

\begin{figure}[!htb]
\centering
\includegraphics[width=.7\columnwidth]{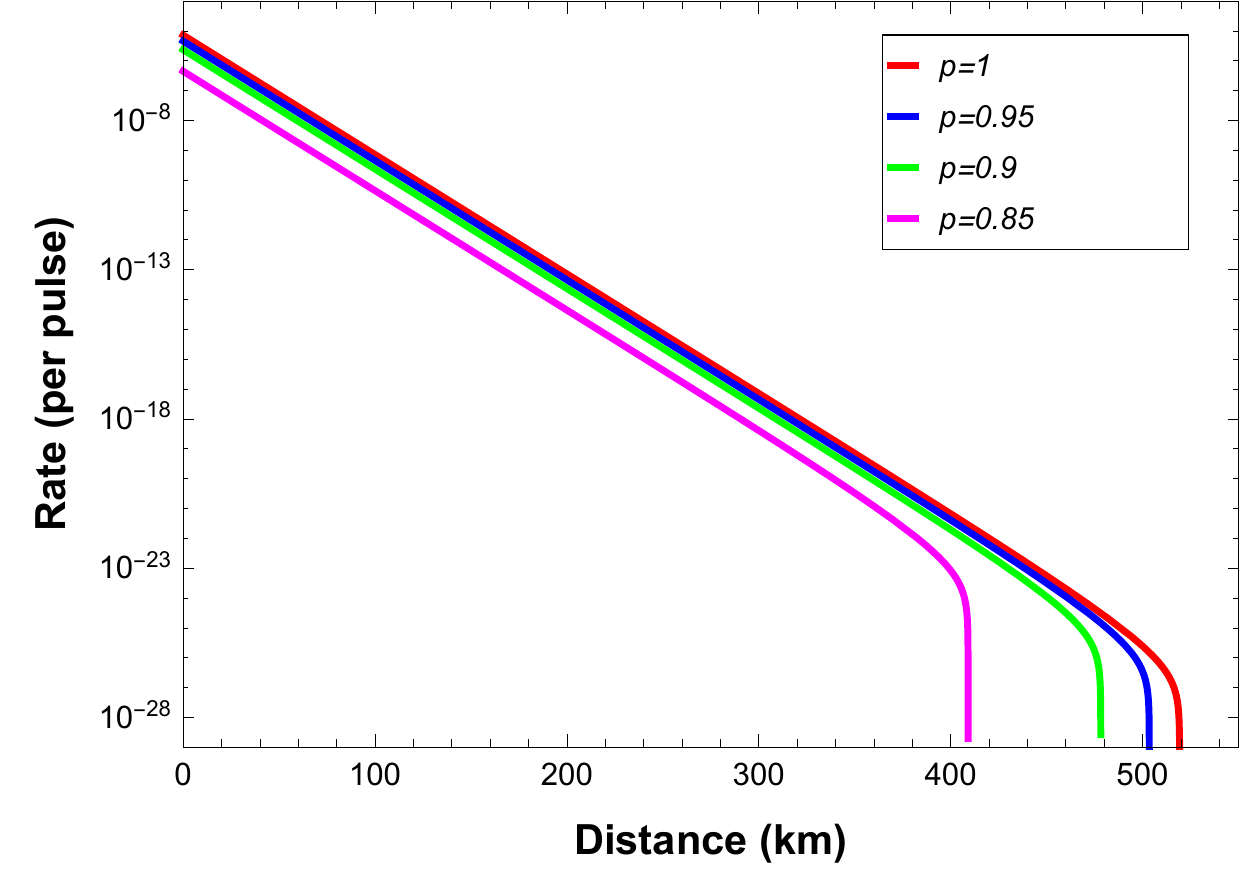}
\caption{\label{fig:key-rate} Lower bound on the secret key rate $R$ versus communication distance between communication users using Werner-like states source.
The red line denotes $p=1$, i.e.,~the perfect GHZ source.
The parameters are chosen according to experiments~\cite{Tang2014a} : the detection efficiency $\eta_{d}=40\%$, the~misalignment-error probability of the system $e_{d}=2\%$, the~dark count rate of the detector $p_{d}=8 \times 10^{-8}$, the~error correction efficiency $f=1.16$, the~intrinsic loss coefficient of the standard telecom fiber channel $\alpha=0.2~\mathrm{dB}/\mathrm{km}$.}
\end{figure}

As an example, we will analyze the secure key rate 
for the $(4,2)$ open-destination MDI-QKD
(see Appendices~\ref{sec:detector-analysis} and~\ref{sec:simulation} for details).
For simplicity, the~single-photon source and the asymptotic approximations are assumed.
We let the BSM setups be located in each user's side, although, in~a realistic experiment, the~BSM setups can be located in anywhere to increase the communication distance.
We suppose that quantum channels are identically depolarizing such that untrusted relays receive the GHZ state in a mixture form~\cite{Chen2004}:
\begin{equation}
\rho = p \ket{GHZ}\bra{GHZ}_{4} + \frac{1-p}{16} \mathbb{I}_{16},
\end{equation}
where $0 \leq p \leq 1$.
We also
assume that all detectors are identical, i.e.,~they have the same dark count rates and the same detection efficiencies.
After numerical simulation, the~lower bound of secure key rates with respective to communication distance between user and central source are shown in~Figure~\ref{fig:key-rate}.

The simulation shows that the secure key rate and the largest communication distance decrease when $p$ decreases.
To implement open-destination MDI-QKD efficiently, good-quality GHZ sources and single-photon sources are necessary.
If such requirements are satisfied, our scheme can tolerate a high loss of more than 500 $\mathrm{km}$ of optical fibers, i.e.,~100 $\mathrm{dB}$, using perfect GHZ source and single-photon source, even when the BSM setups are located in every user's side.
One can double the communication distance by putting the BSM setups in the middle of the users and the GHZ source, which is similar with the case in MDI-QKD~\cite{Lo2012,Braunstein2012}.
For the realistic case where weak coherent pulses are used, our~analysis can be generalized by
considering the decoy state method~\cite{Lo2005, Wang2005} and following the procedures in Refs.~\cite{Curty2014, Xu2014}.

\section{Generalization to The (N,C) Case}

As aforementioned, the~complete analysis has been focused on the $(N,2)$ open-destination MDI-QKD case. Here, we show that the case of two communication users can also generalized to the case of $C$ communication users.
Note that the open-destination feature enables any $C$ users to generate secure keys at the same~time.

Suppose that, in~an $N$-party quantum network with users $1, 2, \cdots, N$, the~communication users are denoted by the subset $\mathcal{C}=\{i_1,i_2,\dots,i_C\}$, where $C=|\mathcal{C}|$.
The auxiliary set denoted by $\mathcal{A}$ consists of auxiliary users, i.e.,~users that assist communication users to generate secure keys, with~$A=|\mathcal{A}| = N - C$ users. According to Equation~(\ref{eq:ghz-reformulating}), for~a general $C$ communication users case, the~GHZ state can be rewritten as
\begin{align}
\ket{GHZ}_{N}
= & \frac{1}{\sqrt{2}}\left[
\ket{00\cdots 0}_{12\cdots C}\bigotimes_{k=C+1\dots N}\frac{\ket{+}_k+\ket{-}_k}{\sqrt{2}}
+ \ket{11\cdots1}_{12\cdots C}\bigotimes_{k=C+1\dots N}\frac{\ket{+}_k-\ket{-}_k}{\sqrt{2}}
\right],\\
= & \left(\frac{1}{\sqrt{2}}\right)^{N-1}\sum_{\chi}
\left(\ket{00\cdots 0}_{12\cdots C}+(-1)^{\sigma_\chi}\ket{11\cdots 1}_{12\cdots C}\right)\ket{\chi}_{C+1\dots N}.
\label{eq:ghz-reformulating-c-user}
\end{align}

Here, $\chi\in\{+,-\}^{N-C}$ is a string of $N-C$ bits with bit value ``$+$'' or ``$-$'' and $\sigma_\chi=0(1)$ if the number of ``$-$'' is even (odd). Intuitively, with~the assistance of $N-C$ auxiliary users, $C$-qubit GHZ states are shared among arbitrary $C$ communication users. Meanwhile, based on the $C$-qubit GHZ state, the~communication users can complete different quantum information tasks with the merit of open destination, such as quantum conference key agreement~\cite{Bose1998,Chen2004,Chen2005a,Fu2015,Zhao2020Phase} and quantum secret sharing~\cite{Hillery1999,Cleve1999,Chen2005,Fu2015}.
In general, we call it the $(N,C)$ open-destination quantum communication task.
When $C=2$, and~the aim is to establish QKD, the~task is reduced to the $(N,2)$ open-destination MDI-QKD network discussed~above.

For instance, in~the general case of $(N,C)$ open-destination quantum conference key agreement, all users prepares and sends BB84 states to their respective untrusted relays.
The central source simultaneously distribute the GHZ state, which is measured together with the state from user on the untrusted relay. When the relays announce successful BSM outcomes and when all auxiliary users announce their prepared states in $X$-basis, the~communication users virtually share a multipartite entangled state, as~the same of the $(N,2)$ case. After~suitable local operations of bit flips, all~communication users share correctly correlated~bits.

By slightly modifying the scheme, the~experimental cost, especially the number of detectors can be reduced significantly.
For instance, when all users announce their preparation basis $X$ for assisting others while keep the bits corresponding to $Z$ basis for distill the key, any $C$ users can share secure keys simultaneously.
This is because their respective sifted keys corresponds to different portions of the raw data.
If one insists on using the conventional two-party QKD and multi-party conference key agreement scheme to realize the same function of the  open-destination scheme under discussion, about $(2^{N}-2)N$ detectors are required. In the open-destination scheme, the~number of detectors is reduced to $4N$, which only increases linearly with the user number $N$.

As an example, we consider the case of $(N,3)$ open-destination quantum conference key agreement. From~Equation~(\ref{eq:ghz-reformulating-c-user}), the~post-selected 3-party GHZ state is $\ket{\phi_{\text{3-party}}^{\pm}}=(\ket{000}\pm\ket{111})/\sqrt{2}$ according to the announcements of the states and the BSM results related with auxiliary users. Meanwhile, as~shown in Table~\ref{tab:map-tab-ghz}, an~equivalent GHZ analyzer among three communication users can be obtained according to the post-selected GHZ state $\ket{\phi_{\text{3-party}}^{\pm}}$ and the BSM results of their corresponding relays. Then, according to the MDI-QCC protocol in Ref.~\cite{Fu2015}, $(N,3)$ open-destination quantum conference key agreement can be directly conducted based on the equivalent GHZ~analyzer.

\begin{table}[!htb]
\caption{The equivalent GHZ analyzer measurement results of three communication users. Here, GHZ$^A$ denotes the post-selected GHZ state from the GHZ source; BSM result 1(2,3) denotes the BSM results of three relays nearby the communication users' side; GHZ analyzer$^C$ denotes the results of corresponding GHZ analyzer among three communication~users.}
\centering
\label{tab:map-tab-ghz}
\begin{tabular}{ccccc}
\toprule
\textbf{GHZ}\boldmath$^A$ & \textbf{BSM Result 1} & \textbf{BSM Result 2} &\textbf{ BSM Result 3} & \textbf{GHZ Analyzer}\boldmath$^C$ \\
\midrule
$\ket{\phi_{\text{3-party}}^{+}}$ ($\ket{\phi_{\text{3-party}}^{-}}$) & $\ket{\psi^+}$ & $\ket{\psi^+}$ & $\ket{\psi^+}$ & $\ket{\phi_{\text{3-party}}^{+}}$ ($\ket{\phi_{\text{3-party}}^{-}}$) \\
$\ket{\phi_{\text{3-party}}^{+}}$ ($\ket{\phi_{\text{3-party}}^{-}}$) & $\ket{\psi^+}$ & $\ket{\psi^+}$ & $\ket{\psi^-}$ & $\ket{\phi_{\text{3-party}}^{-}}$ ($\ket{\phi_{\text{3-party}}^{+}}$) \\
$\ket{\phi_{\text{3-party}}^{+}}$ ($\ket{\phi_{\text{3-party}}^{-}}$) & $\ket{\psi^+}$ & $\ket{\psi^-}$ & $\ket{\psi^+}$ & $\ket{\phi_{\text{3-party}}^{-}}$ ($\ket{\phi_{\text{3-party}}^{+}}$) \\
$\ket{\phi_{\text{3-party}}^{+}}$ ($\ket{\phi_{\text{3-party}}^{-}}$) & $\ket{\psi^+}$ & $\ket{\psi^-}$ & $\ket{\psi^-}$ & $\ket{\phi_{\text{3-party}}^{+}}$ ($\ket{\phi_{\text{3-party}}^{-}}$) \\
$\ket{\phi_{\text{3-party}}^{+}}$ ($\ket{\phi_{\text{3-party}}^{-}}$) & $\ket{\psi^-}$ & $\ket{\psi^+}$ & $\ket{\psi^+}$ & $\ket{\phi_{\text{3-party}}^{-}}$ ($\ket{\phi_{\text{3-party}}^{+}}$) \\
$\ket{\phi_{\text{3-party}}^{+}}$ ($\ket{\phi_{\text{3-party}}^{-}}$) & $\ket{\psi^-}$ & $\ket{\psi^+}$ & $\ket{\psi^-}$ & $\ket{\phi_{\text{3-party}}^{+}}$ ($\ket{\phi_{\text{3-party}}^{-}}$) \\
$\ket{\phi_{\text{3-party}}^{+}}$ ($\ket{\phi_{\text{3-party}}^{-}}$) & $\ket{\psi^-}$ & $\ket{\psi^-}$ & $\ket{\psi^+}$ & $\ket{\phi_{\text{3-party}}^{+}}$ ($\ket{\phi_{\text{3-party}}^{-}}$) \\
$\ket{\phi_{\text{3-party}}^{+}}$ ($\ket{\phi_{\text{3-party}}^{-}}$) & $\ket{\psi^-}$ & $\ket{\psi^-}$ & $\ket{\psi^-}$ & $\ket{\phi_{\text{3-party}}^{-}}$ ($\ket{\phi_{\text{3-party}}^{+}}$) \\
\midrule
\end{tabular}
\end{table}
\vspace{-6pt}
Similar to the open-destination MDI-QKD in Section~(\ref{sec:N_2_case}) of the $(N,2)$ case, the~security of the $(N,3)$ open-destination quantum conference key agreement is also based on the entanglement purification discussion~\cite{lo1999unconditional,shor2000simple,Fu2015}. According to the multi-partite entanglement purification scheme~\cite{Maneva2002}, the~secret key rate can be written as follows~\cite{Chen2004,Fu2015,Zhao2020Phase}:
\begin{equation}
R_3 = Q^Z\{1-f\cdot\max[H(E^Z_{12}),H(E^Z_{13})]-H(E^X)\},
\end{equation}
where $Q^Z$ is the overall gains when three communication users send out quantum states in $Z$ basis, $E^Z_{12}$ ($E^Z_{13}$) is the marginal quantum bit error rate between user $1$ and user $2$ ($3$) in $Z$ basis, $E^X$ is the overall quantum bit error rate in $X$ basis, $f$ is the error correction efficiency, and~\mbox{$H(x) = -x \log_2 (x)-(1-x) \log_2(1-x)$} is the binary Shannon entropy function. $Q^Z$, $E^X$, $E^Z_{12}$, and~$E^Z_{13}$ can be gotten directly from the experimental results. Meanwhile, the~estimation of key rate can be slightly different if the sources of users are weak coherent states~\cite{Gottesman2004}.

\section{Conclusions}\label{sec:conclusion}
As a conclusion, we proposed a flexible and extensible scheme of the $(N,2)$ open-destination MDI-QKD network.
We proved the correctness and security of the protocol, and~derived practical key generation rate formula.
For an illustration, we studied a specific network where two of four users want to distill quantum secure keys.
For the scenario, we presented a polarization-encoding scheme for experimental implementation and offered in detail
a simulation by taking the imperfections in both source and detectors into account. The~simulation results show that the scheme
enjoys a promising structure and performance in real-life~situation.

A significant virtue of our scheme is the security against untrustful relays and all detector side-channel attacks.
Moreover, the~open-destination feature enables any two users to establish MDI-QKD without changing the network structures.
In fact, one can establish MDI-QKD among arbitrary users even after the entangled source have been distributed and all the measurements have been completed.
Furthermore, following the multi-entanglement swapping scheme, the~network can be extended into a large scale by adding shared multi-partite GHZ~states.

We would like to remark that currently the efficiency was relatively low (seen from Figure~\ref{fig:key-rate}).
This can be overcome by taking optimization in network topology, basis selections, and~measurements for both the auxiliary and communication parties, as~well as considering asymmetric loss for various channels, etc., like techniques adopted in Ref.~\cite{Wang2019}.
Any future improvement on distributing multipartite entanglement efficiently and effectively will definitely benefit the proposed scheme and push it forward practical~applications.


\begin{acknowledgments}
We thank Yu-Ao Chen and Qiang Zhang for valuable and enlightening~discussions. 
This research was funded by the Chinese Academy of Science, the~National Fundamental Research Program, the~National Natural Science Foundation of China (Grants No. 11575174, No. 11374287, No. 61125502, No. 11574297, and~No. 61771443), as~well as the Fundamental Research Funds for the Central Universities (WK2340000083).
\end{acknowledgments}




\appendix



\section{Sifting Procedure of The (4,2) Case}\label{sec:sifiting}\label{sec:equivalent-bsm}

\begin{figure}[!htb]
\centering
\includegraphics[width=.75\columnwidth]{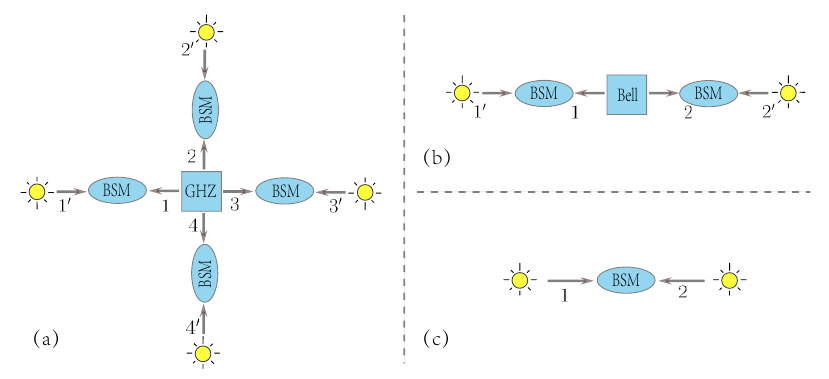}
\caption{
(\textbf{a}) The schematic diagram for the $(4,2)$ open-destination MDI-QKD scheme.
Users $1'$ and $2'$ denote communication users, while users $3'$ and $4'$ denote auxiliary users.
(\textbf{b}) The equivalent topological schematic diagram when users $1'$ and $2'$ communicate with each other.
According to BSM results of relays $3$ and $4$ and quantum states of auxiliary users $3'$ and $4'$, the~GHZ state is projected to a virtual Bell state.
(\textbf{c}) The final equivalent topological schematic diagram that users $1'$~and $2'$ perform MDI-QKD, according to the BSM results and the virtual Bell state.
}
\label{fig:odqkd-security-proof-frame}
\end{figure}

In this section, we describe the sifting procedure of open-destination MDI-QKD in detail for the $(4,2)$ case.
We will show that such scenario can be reduced to the standard MDI-QKD scenario.
The~general case can be proved in a similar way, as~shown in the main text.
The schematic diagram is depicted in Figure~\ref{fig:odqkd-security-proof-frame}a.

We start by writing the GHZ state as
\begin{align}
\ket{GHZ}_{4}
= & \frac{1}{2 \sqrt{2}} [(\ket{00} + \ket{11})(\ket{+ +} + \ket{- -}) \nonumber\\
&  + (\ket{00} - \ket{11})(\ket{+ -} + \ket{- +})].
\label{eq:ghz-reformulating-1}
\end{align}

Up to the announcement of the quantum state of users $3'$ and $4'$, the~BSM(s) of relays $3$ and $4$ on the received quantum state from GHZ source and quantum state from users $3'$($4'$) can be treated as an equivalent projective measurement on the whole GHZ state.
Specifically, if~the relays $3$ and $4$ perform the BSM and obtain equivalent projective measurement results $\ket{00}$ or $\ket{11}$ ($\ket{01}$ or $\ket{10}$), 
the photons received by relays $1$ and $2$ will be projected into state $\ket{\phi^{+}}=(\ket{00} + \ket{11})/\sqrt{2}$ ($\ket{\phi^{-}}=(\ket{00} - \ket{11})\sqrt{2}$) according to Equation~(\ref{eq:ghz-reformulating-1}).
After announcement of the successful BSM results and the quantum states of auxiliary users $3'$ and $4'$, the~projected state received by relays $1$ and $2$ can be determined.
So, one can treat the GHZ source, the~BSM setups of relays $3$ and $4$ and the quantum state of auxiliary user $3'$ and $4'$ as an virtual entanglement source, which outputs different Bell states.
The protocol is thus directly equivalent to MDI-QKD with an entangled source in the middle~\cite{Xu2013a} as illustrated in Figure~\ref{fig:odqkd-security-proof-frame}b.
Since the virtual Bell state with two BSMs along each side can be equivalent to a virtual BSM,
the scheme is finally equivalent to implement MDI-QKD between users $1'$ and $2'$ as showed in Figure~\ref{fig:odqkd-security-proof-frame}c.
Therefore, in~an honest run, the~protocol is reduced to the honest standard MDI-QKD scenario, and~the parties will end up with sharing a common~key.

Firstly, notice that the projection measurement of two systems onto one Bell state can be viewed as a POVM (positive operator valued measure) 
on one system if one knows the state of the other system.
For example, as~shown in Figure~\ref{fig:odqkd-security-proof-frame}a,
a successful BSM result of $\ket{\psi^{-}}$ of the relay $3$ with auxiliary photons from auxiliary $3'$ in the state $\ket{\alpha}_3^\prime$
can be viewed as a POVM ${\rm tr}_{3^\prime}\left[\ket{\psi^-}\bra{\psi^-}_{33^\prime}\ket{\alpha}\bra{\alpha}_{3^\prime} \right]$ on the state $3$.
In the open-destination scheme, we have $\ket{\alpha}\in\{\ket{+},\ket{-}\}$ and the BSM results $\{\ket{\psi^{+}},\ket{\psi^{-}}\}$.
The correspondence between the POVM on the system $k$ and the untrusted relay announces a successful BSM together with auxiliary state
are listed in Table~\ref{tab:map-tab-projection}.

\begin{table}[!htb]
\caption{The correspondence between the POVM 
on state labeled $k$ and the BSM result labeled by $kk^\prime$ with auxiliary state labeled by $k^\prime$.}
\centering
\label{tab:map-tab-projection}
\begin{tabular}{ccc}
\toprule
\textbf{State of System} \boldmath$k^\prime$ & \textbf{BSM Result on Systems} \boldmath$kk^\prime$ & \textbf{POVM on System} \boldmath $k$ \\
\midrule
$\ket{+}$ & $\ket{\psi^-}$ & $\ket{-}\bra{-}/2$\\
$\ket{-}$ & $\ket{\psi^-}$ & $\ket{+}\bra{+}/2$\\
$\ket{+}$ & $\ket{\psi^+}$ & $\ket{+}\bra{+}/2$\\
$\ket{-}$ & $\ket{\psi^+}$ & $\ket{-}\bra{-}/2$\\
\midrule
\end{tabular}
\end{table}
\unskip
\vspace{6pt}
Secondly, when the two auxiliary users prepare $X$-basis photons and the corresponding relays get successful BSM results,
according to Table~\ref{tab:map-tab-projection}, the~total GHZ state collapses into one of the maximally entangled states $\ket{\phi^{\pm}}=\frac{1}{\sqrt{2}}(\ket{H H} \pm \ket{V V})$ at the side of two communication users.

Thirdly, at~the sides of the two communication users, according to the post-selected Bell state $|\phi^\pm\rangle$ and the BSM results of their corresponding relays, a~BSM between two communication users can be obtained.
Such correspondence is listed in Table~\ref{tab:map-tab-bsm}.

\begin{table}[!htb]
\caption{The equivalent BSM results of two communication users. Here, Bell$^A$ denotes the post-selected Bell state from the GHZ source; BSM result 1(2) denotes the BSM results of the two relays nearby the communication users' side; BSM$^C$ denotes the results of corresponding BSM between two communication~users.}
\centering
\label{tab:map-tab-bsm}
\begin{tabular}{cccc}
\toprule
\textbf{Bell}\boldmath$^A$ & \textbf{BSM Result 1} & \textbf{BSM Result 2} & \textbf{BSM}\boldmath$^C$ \\
\midrule
$\ket{\phi^{+}}$ & $\ket{\psi^+}$ & $\ket{\psi^+}$ & $\ket{\phi^+}$ \\
$\ket{\phi^{+}}$ & $\ket{\psi^+}$ & $\ket{\psi^-}$ & $\ket{\phi^-}$ \\
$\ket{\phi^{+}}$ & $\ket{\psi^-}$ & $\ket{\psi^+}$ & $\ket{\phi^-}$ \\
$\ket{\phi^{+}}$ & $\ket{\psi^-}$ & $\ket{\psi^-}$ & $\ket{\phi^+}$ \\
$\ket{\phi^{-}}$ & $\ket{\psi^+}$ & $\ket{\psi^+}$ & $\ket{\phi^-}$ \\
$\ket{\phi^{-}}$ & $\ket{\psi^+}$ & $\ket{\psi^-}$ & $\ket{\phi^+}$ \\
$\ket{\phi^{-}}$ & $\ket{\psi^-}$ & $\ket{\psi^+}$ & $\ket{\phi^+}$ \\
$\ket{\phi^{-}}$ & $\ket{\psi^-}$ & $\ket{\psi^-}$ & $\ket{\phi^-}$ \\
\midrule
\end{tabular}
\end{table}
\vspace{-6pt}

Finally, as~shown in Table~\ref{tab:flip}, according to the final equivalent BSM result and the preparation bases, one of the communication users apply a bit flip or not such that their keys can be correlated.
In~fact, only when both communication users select $X$ basis and the final equivalent BSM result is $\ket{\phi^-}$, one of them needs to apply a bit flip.
After many rounds, they obtain enough raw key bits that can be used in the following data post-processing~process.

\begin{table}[!htb]
\caption{Flip table according to the preparation bases and the equivalent BSM result at communication users~side.}
\centering
\label{tab:flip}
\begin{tabular}{ccc}
\toprule
\textbf{ Basis} & \boldmath$\ket{\phi^+}$ & \boldmath$\ket{\phi^-}$\\
\midrule
$Z$-basis & No Flip & No Flip \\
$X$-basis & No Flip & Flip \\
\midrule
\end{tabular}
\end{table}
\unskip

\section{Detector~Analysis}\label{sec:detector-analysis}

\begin{figure}[!htb]
\centering
\includegraphics[width=.5\columnwidth]{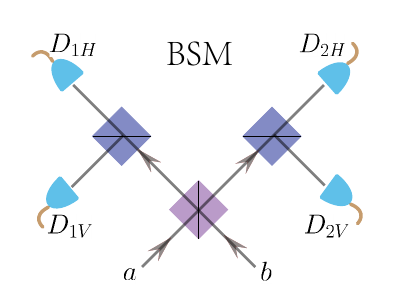}
\caption{The BSM setup with polarization encoding. BS denotes beam splitter, PBS denotes polarization beam splitter, and~$H$ and $V$ denote, respectively, horizontal and vertical linear polarizations, and~$D_{1H}$, $D_{2H}$, $D_{1V}$, $D_{2V}$ denote single-photon detectors.
A click in $D_{1H}$ and $D_{2V}$, or~in $D_{1V}$ and $D_{2H}$, indicates a projection into the Bell state $\ket{\psi^{-}}=(\ket{HV}-\ket{VH})/\sqrt{2}$, and~a click in $D_{1H}$ and $D_{1V}$, or~in $D_{2H}$ and $D_{2V}$, indicates a projection into the Bell state $\ket{\psi^{+}}=(\ket{HV}+\ket{VH})/\sqrt{2}$.}
\label{fig:bsm-setup}
\end{figure}

Since the BSM with the auxiliary photon is equivalent to an probabilistic projective measurement, one can use an equivalent detector to replace the BSM device with the corresponding light source in the key rate analysis.
Here, we develop a method to derive the equivalent detector parameters, i.e.,~the detection efficiency and the dark count of the equivalent detector.
We use the BSM setup with polarization encoding as illustrated in Figure~\ref{fig:bsm-setup}.

In $H/V$ basis, suppose that Alice and Bob encode the same polarization states; then, the~state becomes as follows after the BS:
\begin{equation}
a^{\dagger}_{H} b^{\dagger}_{H} \ket{vac} \rightarrow (a^{\dagger 2}_{1H} - a^{\dagger 2}_{2H})\ket{vac},
\end{equation}
where $a^{\dagger}$ ($b^{\dagger}$) denotes creation operators, and~$\ket{vac}$ denotes vacuum state.
The probability of the successful BSM when the input states are $\ket{H}$ and $\ket{H}$, is given by
\begin{equation}
P_{H H} = 2 p_d (1-p_d)^2 (1-(1-p_d)(1-\eta_d)^2),
\end{equation}
where $\eta_d$ is the detection efficiency, and~$p_d$ is the dark count.
Suppose that Alice and Bob encode different polarization state; then, after~the BS, the~state becomes as follows:
\begin{equation}
\begin{aligned}
a^{\dagger}_{H} b^{\dagger}_{V} \ket{vac} \rightarrow & (a^{\dagger}_{1H}a^{\dagger}_{1V} -  a^{\dagger}_{2H}a^{\dagger}_{2V}) \ket{vac} \\
& + (a^{\dagger}_{2H}a^{\dagger}_{1V} - a^{\dagger}_{1H}a^{\dagger}_{2V})\ket{vac}.
\end{aligned}
\end{equation}

The probability of the successful BSM when the input states are $\ket{H}$ and $\ket{V}$ is given by
\begin{equation}
P_{H V}=  (1-p_d)^2 (1-(1-p_d)(1-\eta_d))^2.
\end{equation}

Thus, the~equivalent detection probability when the input state is $\ket{H}$ is given by
\begin{equation}
\begin{aligned}
\eta^{'}_{H} = & \frac{1}{2} (1-p_d)^2 [2p_d(1-(1-p_d)(1-\eta_d)^2)\\
& +(1-(1-p_d)(1-\eta_d))^2].
\end{aligned}
\end{equation}

Due to symmetry, the~equivalent detection probability when the input state is $\ket{V}$ has the same form with the case that the input state is $\ket{H}$, i.e.,~one has $\eta^{'}_{V}=\eta^{'}_{H}$.
Similarly, by~using the transformation relation under $\{+,-\}$ basis
\begin{equation}
\begin{aligned}
a^{\dagger}_{+} b^{\dagger}_{+} \ket{vac} &\rightarrow (a^{\dagger}_{1H}a^{\dagger}_{1V} - a^{\dagger}_{2H}a^{\dagger}_{2V}) \ket{vac} \\
a^{\dagger}_{+} b^{\dagger}_{-} \ket{vac} &\rightarrow (a^{\dagger}_{1H}a^{\dagger}_{2V} - a^{\dagger}_{1V}a^{\dagger}_{2H}) \ket{vac},
\end{aligned}
\end{equation}
one can ontain the equivalent detection probability when the input state is $\ket{+}$ as follows:
\begin{equation}
\begin{aligned}
\eta^{'}_{+} = & (1-p_d)^2 (1-(1-p_d)(1-\eta_d))^2.
\end{aligned}
\end{equation}

Due to symmetry, one has $\eta^{'}_{-}=\eta^{'}_{+}$.

We consider practical experimental parameters, which are listed in Table~\ref{tab:parameters}. For~the experimental parameters, one arrives at
\begin{equation}
\eta_d^{'Z} = 0.08, \quad \eta_d^{'X} = 0.16,
\end{equation}
where $\eta_d^{'Z}$ denotes the equivalent detection efficiency for $H/V$ basis, i.e.,~$Z$ basis, and~$\eta_d^{'X}$ denotes the equivalent detection efficiency for $+/-$ basis, i.e.,~$X$ basis.

\begin{table}[!htb]
\caption{\label{tab:parameters}
List of experimental parameters used for simulation.
$\eta_{d}$ is the detection efficiency; $e_{d}$ is the misalignment-error probability of the system; $p_{d}$ is the dark count rate of the detector; $f$ is error correction efficiency; $\alpha$ is the intrinsic loss coefficient of the standard telecom fiber channel.}
\centering
\begin{tabular}{ccccc}
\toprule
\boldmath$\eta_d$ & \boldmath$e_d$ & \boldmath$p_d$ & \boldmath$f$ & \boldmath$\alpha(\mathrm{dB}/\mathrm{km})$\\
\midrule
$40\%$ & $2\%$ & $8\times10^{-8}$ & $1.16$ & $0.2$\\
\midrule
\end{tabular}
\end{table}
\vspace{-6pt}

To calculate the parameters for equivalent dark count, one should consider the case in which there was no incoming photon.
Suppose the local photon being $\ket{H}$, and~the incoming photon being vacumm state, the~states become as follows after the BS:
\begin{equation}
b_{H}^{\dagger}	\ket{vac} \rightarrow \frac{i}{\sqrt{2}} a_{1H}^{\dagger} + \frac{1}{\sqrt{2}} a_{2H}^{\dagger},
\end{equation}
where $b_{H}^{\dagger}$ denotes the creation operator of local photon.
So, one can get the probability of the successful BSM as follows:
\begin{equation}
P_{H} =  2 p_d (1-p_d)^2	 \eta_d.
\end{equation}

Due to symmetry, one has that $P_{+}=P_{-}=P_{V}=P_{H}$. Here, $P_{x}$ denotes the probability of the successful BSM result when the local photon is $\ket{x}$ and there is no incoming photon.
So, one can get the equivalent dark count as
\begin{equation}
p_d^{'} =  2 p_d (1-p_d)^2 \eta_d .
\end{equation}

For the experimental parameters given in Table~\ref{tab:parameters}, one arrives at
\begin{equation}
p_d^{'} = 6.4 \times 10^{-8}.
\end{equation}

Finally, one can achieve the parameters for the equivalent detectors shown in Table~\ref{tab:equivalent-parameters}.

\begin{table}[!htb]
\caption{\label{tab:equivalent-parameters}
List of the parameters for the equivalent detectors.
$\eta_d^{'Z}$ ($\eta_d^{'X}$ ) denotes the equivalent detection efficiency for $Z$ ($X$) basis, and~$p_d^{'}$ denotes the equivalent dark count.
}
\centering
\begin{tabular}{ccc}
\toprule
\boldmath$\eta_d^{' Z}$ & \boldmath$\eta_d^{' X}$ & \boldmath$p_d^{'}$ \\
\midrule
$8\%$ & $16\%$ & $6.4\times10^{-8}$ \\
\midrule
\end{tabular}
\end{table}
\unskip

\section{Simulation for (4,2)-Scenario}\label{sec:simulation}

\begin{figure}[!htb]
\centering
\includegraphics[width=.9\columnwidth]{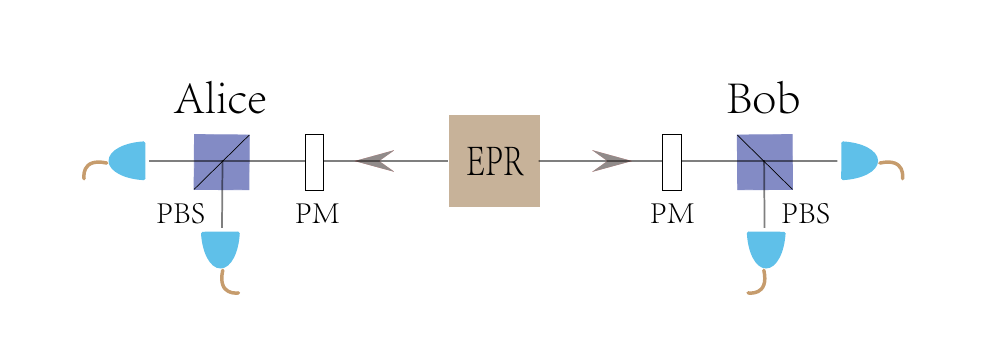}
\caption{Equivalent setup for Alice and Bob when tracing the BSM results of the auxiliary users. PBS~denotes polarization beam splitter, PM denotes polarization modulator, and~EPR 
denotes~EPR~source.}
\label{fig:equivalent-setup}
\end{figure}

For simulation purposes, one can assume practically that the source has the form of Werner-like~states
\begin{equation}
\rho = p \ket{GHZ}\bra{GHZ}_{4} + \frac{1-p}{16} \mathbb{I},
\end{equation}
in which $\ket{GHZ}_{4} = (\ket{H H H H}+\ket{V V V V})/\sqrt{2}$ is the $4$-partite GHZ states, $\mathbb{I}/16$ is the $4$-partite maximal mixed states, and~$0 \leq p \leq 1$.
As proven in the previous section, according to the measurement results of auxiliary side, the~photons received by communication side will be projected into different Bell states.
Here, we consider the case in which auxiliary side get the $\ket{+}\otimes\ket{+}$ results, due to the symmetry.
When auxiliary side get the $\ket{+}\otimes\ket{+}$ result, the~particles received by communication side will collapse into
\begin{equation}
\rho_{\text{AB}} = p \ket{\phi^+}\bra{\phi^+} + \frac{1-p}{4} \mathbb{I},
\end{equation}
where $\phi^+ = (\ket{H H}+\ket{V V})/\sqrt{2}$ is one of the Bell states.
So, it is equivalent with the case in which the two communication users (denoted by Alice and Bob) perform an entanglement-based QKD using the two-qubit Werner states $\rho_{AB}$ as a source and the equivalent detectors as detection device, as~illustrated in Figure~\ref{fig:equivalent-setup}, from~the perspective of key rate~analysis.

Taking these imperfections of the source and detectors into account, the~key generation rate in a realistic setup will be given by
\begin{equation}
R=Q_{11}^{ZZ}(1 - H(e_{11}^{XX}))-Q_{\mu\nu}^{ZZ}\cdot f \cdot H(E_{\mu\nu}^{ZZ}).
\end{equation}

In the following, we discuss how one can derive each quantity in this key rate formula, i.e.,~$Q_{11}^{ZZ}$, $e_{11}^{XX}$, $Q_{\mu\nu}^{ZZ}$, and~$E_{\mu\nu}^{ZZ}$.\vspace{-6pt}

\textit{Yield.} Denote the yield of single-photon pair as $Y_{11}$, i.e.,~the conditional probability of a coincidence detection event given that the entanglement source emits an single-photon pair. Then, $Y_{11}$ is given by
\begin{equation}
Y_{11} = [1- (1-Y_{0A})(1-\eta_A)][1- (1-Y_{0B})(1-\eta_B)], \label{eq:Y1}
\end{equation}
where $Y_{0A} = Y_{0B} = p_{d}^{'}$ are the background count rates on Alice's and Bob's sides in the $Z$ basis, and~$\eta_A=\eta_B=\eta_{d}^{' Z}\times10^{-\alpha L/20}$ denotes the total detection efficiency considering the channel loss. Equation~(\ref{eq:Y1}) is also applicable to the $X$ basis.
Then, the~gain of the single photon part and the overall gain are given by
\begin{equation}
Q_{\mu\nu}^{ZZ} = Q_{11}^{ZZ} = Y_{11}.
\end{equation}

\textit{Error Rate.}
The error rate of single-photon pair in the $X$ basis $e_{11}^{XX}$ has three main contributions taking some imperfections into account:
(i) \textit{The imperfections of entanglement source}, i.e.,~the maximal mixed states component, which brings $50\%$ error rate $e_0=1/2$;
(ii) \textit{Background counts}, which are random noises $e_0=1/2$;
(iii) \textit{Intrinsic detector error $e_d$}, which characterizes the alignment and stability of the optical system.
So, the~error rate of single-photon pair $e_{11}^{XX}$ is given as follows:
\begin{equation}
e_{11}^{XX} Y_{11} =  p e_0 (Y_{11} - \eta_A \eta_B) + p e_d \eta_A \eta_B +  (1-p) e_0 Y_{11},
\end{equation}
where the first item comes from background counts, the~second term comes from intrinsic errors, and~the third term comes from the mixed part of the source.
So, one achieves the error rate of single-photon pair $e_{11}^{XX}$ as follows:
\begin{equation}
e_{11}^{XX} = e_0 - \frac{ p  \eta_A \eta_B (e_0 - e_d)}{Y_{11}}.
\end{equation}

Similarly, the~error rate in the $Z$ basis is given by
\begin{equation}
E_{\mu\nu}^{ZZ} = e_0 - \frac{ p  \eta_A \eta_B (e_0 - e_d)}{Y_{11}}.
\end{equation}


\end{document}